\documentclass[10pt]{article}
\DeclareUnicodeCharacter{0301}{\'{e}}
\usepackage[utf8]{inputenc}
\usepackage[T1]{fontenc}
\usepackage{authblk}
\usepackage[svgnames,table]{xcolor} 
\usepackage{hyperref}
\usepackage{amsmath}
\usepackage{amssymb} 
\usepackage[utf8]{inputenc}
\usepackage{graphicx}
\usepackage{subcaption}
\usepackage{float} 
\usepackage{caption}
\usepackage[backend=biber,style=numeric,sorting=none]{biblatex} 
\usepackage{geometry}
\usepackage{tabularx,tabulary}
\usepackage{tikz}
\usetikzlibrary{shapes.geometric, arrows}

\tikzstyle{startstop} = [rectangle, rounded corners, 
minimum width=3cm, 
minimum height=1cm,
text centered, 
draw=black, 
fill=gray!30]

\tikzstyle{io} = [trapezium, 
trapezium stretches=true, % A later addition
trapezium left angle=70, 
trapezium right angle=110, 
minimum width=3cm, 
minimum height=1cm, text centered, 
draw=black, fill=gray!30]

\tikzstyle{process} = [rectangle, 
minimum width=3cm, 
minimum height=1cm, 
text centered, 
text width=3cm, 
draw=black, 
fill=gray!30]

\tikzstyle{decision} = [diamond, 
minimum width=3cm, 
minimum height=1cm, 
text centered, 
draw=black, 
fill=gray!30]
\tikzstyle{arrow} = [thick,->,>=stealth]

\title{Random matrix theory and nested clustered portfolios on Mexican markets}

\date{}
\author[1,2,*]{Andr\'{e}s Garc\'{i}a-Medina}
\author[1]{Benito Rodríguez-Camejo }
\affil[1]{Centro de Investigación en Matemáticas, Unidad Monterrey,  Av. Alianza Centro  502, PIIT 66628, Apodaca, Nuevo León, México}
\affil[2]{Consejo Nacional de Humanidades, Ciencias y Tecnologías, Av. Insurgentes Sur 1582, Col. Crédito Constructor 03940, Ciudad de México, México}

\affil[*]{Corresponding author: andres.garcia@cimat.mx}

\begin{document}

\flushbottom
\maketitle

\thispagestyle{empty}

\begin{abstract}
This work aims to deal with the optimal allocation instability problem of Markowitz's modern portfolio theory in high dimensionality.
We propose a combined strategy that considers covariance matrix estimators from Random Matrix Theory~(RMT) and the machine learning allocation methodology known as Nested Clustered Optimization~(NCO).
The latter methodology is modified and reformulated in terms of the spectral clustering algorithm and Minimum Spanning Tree~(MST) to solve internal problems inherent to the original proposal.
Markowitz's classical mean-variance allocation and the modified NCO machine learning approach are tested on financial instruments listed on the Mexican Stock Exchange~(BMV) in a moving window analysis from 2018 to 2022.
The modified NCO algorithm achieves stable allocations by incorporating RMT covariance estimators.
In particular, the allocation weights are positive, and their absolute value adds up to the total capital without considering explicit restrictions in the formulation.
Our results suggest that can be avoided the risky \emph{short position} investment strategy by means of RMT inference and statistical learning techniques.
%%%

\end{abstract}


\section{Introduction}

Technology has had many advances in the information age, including the increasing amount of accessible financial data. 
In this scenario, it is not uncommon to have many variables, even more than transaction days, where each variable or asset can be considered an extra dimension, meaning that multivariate data exists in a high-dimensional space.
Thus, this enormous volume of data and the new technologies to process it have opened the doors for applications in high-dimensionality multivariate statistics.

One of the fundamental characteristics of multivariate data is its covariance matrix. The covariance elements can be interpreted as the tendency of variables to vary together.
The estimation of a covariance matrix or its inverse, the precision matrix, is a topic of great relevance in real-life situations. One of these cases is the optimization of portfolios, where the portfolio risk is modeled based on the covariance matrix estimation. However, when the dimension is comparable to the sample size, the empirical covariance estimator is ill-conditioned and contains estimation errors. Hence, it is suggested to use high-dimensionality estimation methods framed on Random Matrix Theory~(RMT) and work with the spectrum of the covariance matrix.
%%% REVIEW FROM HERE...
RMT is a new type of statistical mechanics where instead of having an ensemble of states governed by the same Hamiltonian, one has an ensemble governed by the same symmetry.
The origins of random matrices go back to the physics of the 1950s. At that time, Eugene Wigner proposed a purely statistical description for studying the energy levels of the uranium nucleus~\cite{mehta2004random}. Wigner hypothesized that the statistical behavior of energy levels could be well modeled by the eigenvalues of a random matrix. His model replaces the Hamiltonian matrix of the system with a finite but large random matrix called the Wigner matrix. This matrix is symmetric (or Hermitian in the complex case), whose entries are independent and identically distributed with zero mean and a finite variance.
In 1962 Dyson \cite{dyson1962threefold} extended Wigner's ideas, showing that physically reasonable symmetry assumptions can be represented through Gaussian ensembles.
The probability density of finding a particular matrix within one of these
ensembles are invariant under Orthogonal, Unitary, and Symplectic transformations of the Hamiltonian and are known as Gaussian Orthogonal Ensemble (GOE), Gaussian Unitary Ensemble (GUE), and Gaussian Symplectic Ensemble (GSE), respectively.

On the other hand, the Wishart \cite{wishart1928generalised} distribution describes the covariance matrix of multivariate Gaussian data. In the context of RMT, the covariance matrices are often called members of the Wishart Ortoghonal Ensemble (WOE). The Wishart distribution can be seen as generalizing the relationship between the univariate normal and the chi-square distribution~\cite{pena2002analisis}.

We are interested in applying the RMT techniques to improve the covariance matrix estimation involved in the portfolio optimization problem.
The modern portfolio theory of Markowitz~\cite{markowitz1952} formulates the asset allocation problem as a quadratic optimization problem.
A weakness of Markowitz's portfolio theory is that if the non-diagonal elements of the covariance matrix are not negligible, the condition number will be high. Therefore the allocation solutions to the optimization problem will be unstable. In other words, the Markowitz theory intends to diversify the allocation weights of an investment portfolio. However, the instability increases with the number of assets.
Hence, the more we need to use the Markowitz strategy, the less reliable the solution is. This paradox is known as \emph{Markowitz's curse}.
The Nested Clustered Optimization (NCO) algorithm is a strategy to deal with Markowitz's curse of highly correlated portfolio assets proposed by de Prado~\cite{de2020machine}.
The procedure is based on Markowitz's mean-variance approach and tries to spread the instability of the portfolio through its constituent blocks. A clustering method is applied to this aim, and the optimal intracluster and intercluster allocations are computed to obtain a set of stable weights.

The original proposal of NCO~\cite{lopez2016robust} algorithm uses the k-means clustered algorithm~\cite{johnson2002applied} and the silhouette~\cite{rousseeuw1987silhouettes} method to determine the composition and number of groups of the dissimilarity matrix associated to the covariance matrix.
However, the k-means method should not be used on the dissimilarity matrix but on the data matrix.
Nevertheless, working with the dissimilarity matrix rather than the data matrix in portfolio optimization is more informative to account for the interactions between financial assets. Hence,  we reformulate the NCO in terms of the spectral clustering algorithim~\cite{von2007tutorial}, and the minimum spanning tree~(MST)~\cite{west2001introduction} to avoid the methodological error of using the k-means algorithm.
The MST has been initially proposed in the econophysics community to explain the heterogeneous structure of financial markets~\cite{mantegna1999hierarchical}. In comparison, spectral clustering is a machine-learning technique with roots in graph theory~\cite{donath1973lower,fiedler1973algebraic}.

Thus, the subject of this research stems from an interest in strengthening estimates and calculations in the world of investment portfolios. 
We would like to know if applying estimation techniques from RMT can reduce the underlying risk of investment portfolios. In addition, we are interested in whether these improvements are reflected in the portfolio optimization methods in the sense of an increase in the diversification of the portfolio and a better allocation of assets.
To fulfill this purpose~\footnote{The purpose of this paper is inspired by the Master thesis of one of the authors~\cite{Benito2022}}, we analyze the effect of applying high-dimensionality estimators of the covariance matrix under the two discussed portfolio optimization strategies. In particular, it is applied the optimal linear estimator~\cite{ledoit2004well} and an estimator based on the limited Tracy-Widom distribution~\cite{tracy1994level}, both based on RMT. The portfolios are built using the Markowitz mean-variance approach and our allocation variation of NCO.
The strategies are tested and contextualized in a set of assets listed on the Mexican Stock Exchange~(BMV) from 2018 to the first half of 2022.

The paper is organized as follows. Section 2 introduces the RMT estimators formulated since the recent statistical inference approach. 
Section 3 presents the main elements of portfolio theory and the allocation methodologies of Markowitz and NCO. Section 4 describes the financial data. Section 5 shows and analyze the results of the different estimators and allocation strategies on the BMV. Finally, section 6 features the conclusions and discusses future work.

\section{Random Matrix Theory Estimators}

To start with, consider a data matrix $X$ of dimension $p\times n$ matrix with i.i.d. Gaussian vector entries  $X_{i}\sim N_p(0,\mathbf{\Sigma}), i=1,\dots,n$, i.e., with zero mean and covariance matrix $\mathbf{\Sigma}$.
The joint distribution $f(w_{11},\dots,w_{pp})$ of the $\frac{1}{2}p(p+1)$ distinct elements of  $\mathbf{W = XX}^{T}$ is denoted as $W_p(n,\mathbf{\Sigma})$, and given by the Wishart distribution~\cite{pena2002analisis,wishart1928generalised}
    \begin{equation}
        f(w_{11},\dots,w_{pp}) = c |\mathbf{\Sigma}|^{-1/2}|\mathbf{W}|^{(n-p-1)/2} \exp{\{ -\frac{1}{2} Tr \mathbf{W}\}},
    \end{equation}
where $c$ is a normalization constant and is assumed $\mathbf{\Sigma}$ to be positive definite with $n\geq p$.

In RMT, the $\mathbf{W}$ matrix is said to be a member of the WOE because its probability within the ensemble is invariant under orthogonal transformations.
An important results of this ensemble under the assumption $\mathbf{\Sigma=I}$ is the as $p,n\rightarrow\infty$, such that $\frac{p}{n}\rightarrow q \in (0,\infty)$, the eigenvalue distribution of the scaled $\mathbf{W}/n$ converges (almost surely) to the Marchenko-Pastur law\cite{marvcenko1967distribution}
\begin{equation}
    \rho(\lambda) = \frac{\sqrt{(\lambda_{+}-\lambda)(\lambda-\lambda_{-})}}{2\pi q \lambda},\quad \lambda_{\pm} = (1\pm\sqrt{q})^2
\end{equation}

It is possible to place the Wishart distribution in the language of statistical inference and calculate confidence intervals to accept or reject the Marchenko-Pastur law.
The confidence intervals are constructed considering the probability of finding the eigenvalue $\lambda_1$ of $\mathbf{W}$ larger than an arbitrary upper bound $M$, given that $\mathbf{W}$ follows the white Wishart distribution $ W_p(0,\mathbf{I})$
$\lambda_1$~\cite{Iain2006}
  \begin{equation}
    P\{\lambda_1 > M: \mathbf{W} \sim W_p(n,\mathbf{I})\}.
  \end{equation}

The statistical inference in random matrices is possible thanks to the theoretical works of Craig Tracy and Harold Widom~\cite{tracy1994level, tracy1996orthogonal}, who found that the probability distribution of the largest eigenvalue of random matrices belonging to several ensembles with symmetry properties denoted by $\beta$,  closely approximates to
  \begin{equation}
    P\{ n\hat{\lambda}_1 \leq \mu_{np} + \sigma_{np} s | \mathbf{A}\} \rightarrow F_{\beta}(s)
    \label{tracy_widom}
   \end{equation}
with appropriate centering $\mu_{np}$ and scaling  $\sigma_{np}$ parameters depending on the structure and invariance properties of the arbitrary random matrix $\mathbf{A}$.
This result is known today as the Tracy-Widom law in the context of random matrices and is valid even for more general distributions than Gaussian under certain assumptions. The distribution $F_{\beta}(s)$ is found by solving
\begin{equation}
  \begin{split}
   F_1(s) &= \sqrt{F_2(s) \exp\left( -\int_s^{\infty} q(x) dx \right)}\\
   F_2(s) &= \exp\left( -\int_s^\infty (x-s)^2 q(x) dx\right),
  \end{split}
  \label{TracyEqs}
\end{equation}
which are in terms of the solution $q$ of the non-linear differential equations of 2nd order
\begin{equation}
    \begin{split}
    q'' &= sq + 2q^3\\
    q(s) & \sim A_i(s)\quad\text{when}\quad s\rightarrow \infty.
    \end{split}
\end{equation}
These equations are known as the classical Painlevé type II equations.

Particularly, the members of the WOE reach a  convergence ratio of order $\mathcal{O}(p^{-1/3})$ by the parameters~\cite{johnstone2001distribution}
 \begin{equation}
  \begin{split}
  \mu_{np} &=  (\sqrt{n-1}+\sqrt{p})^2\\
  \sigma_{np} &= (\sqrt{n-1}+\sqrt{p})(1/\sqrt{n-1}+1/\sqrt{p}).
  \end{split}
 \end{equation}
Nevertheless, modifying the scaling and centering parameter
leads to an error of $\mathcal{O}(p^{-2/3})$~\cite{el2006rate}.

The RMT machinery reformulated from the statistical inference point of view enables to test of the null hypothesis of having a covariance matrix coming from the Wishart distribution $H_0: W_p(n,\mathbf{I})$ against the alternative hypothesis  $H_A: W_p(n,\mathbf{\Sigma})$.
Here, we propose to use the Tracy-Widom law as a covariance estimator.
The idea is to reconstruct the covariance matrix with the largest eigenvalues $r$ that violate the null hypothesis $H_0$, leaving the empirical eigenvectors intact. This estimator can be considered the statistical version of the clipping~\cite{laloux1999noise,plerou2002random} technique. Here, the estimated eigenvalues $\xi^{TW}$ are given by
\begin{equation}
     \xi^{TW} = \left\{
                \begin{array}{ll}
		         \bar{\lambda}      & if \quad \text{$H_0$ is true}\\
	        	 \lambda_k    & \text{otherwise}
                \end{array}
                \right.
                \label{RMT}
\end{equation}
where $\bar{\lambda}$ represents the eigenvalues average that satisfy $H_0$. Then, the estimated correlation matrix is given by
    \begin{equation}
        \mathbf{\Xi}^{TW} = \sum_{i=1}^{p} \xi_i^{TW} v_i v_i',
    \end{equation}
where $v_i$ are the eigenvectors of the sample or empirical covariance matrix $\mathbf{E}$.

An estimator that has roots in RMT but has been developed within the area of mathematical statistics with a Bayesian approach is the one proposed by Ledoit and Wolf~\cite{ledoit2004well}.
They propose dealing with the ill-conditioned problem of estimating a covariance matrix when the number of variables $p$ is of the order of observations $n$ through the convex linear combination of the sample covariance matrix with the identity matrix, which is known as \emph{linear shrinkage}
and is expressed as
\begin{equation}
            \mathbf{\Xi}^{linear} = \hat{\alpha} \mathbf{I} + (1-\hat{\alpha})\mathbf{E},
\end{equation}
where $\mathbf{I}$ denotes the identity matrix and $\hat{\alpha}$ is the optimal parameter that shrinks $\mathbf{E}$ to $\mathbf{I}$. In this way, the estimator is the weighted average of the empirical covariance matrix and the matrix where all the variances are the same, whereas the covariances are equal to zero. The optimal value of $\hat{\alpha}$ is found to be asymptotically approximated by
  $\hat{\alpha}$~with respect to a quadratic loss function
\begin{equation}
     \hat{\alpha} = \frac{min\{\frac{1}{n} \sum_{i=1}^n|| x_i x_i' - \mathbf{E} ||_{F}^2, ||\mathbf{E} - \mathbf{I}||_{F}^2\}}{||\mathbf{E} - \mathbf{I}||_{F}^2},
     \label{OptimalLS}
\end{equation}
where $|| \cdot||_{F}$ represents the Frobenius norm and $x_i$ is the i-th column of $\mathbf{E}$.
An advantage of this estimation is that it is not required to assume a specific distribution. It is only necessary to ensure that the data $\mathbf{X}$ have a finite fourth moment.

The last estimator is the sample correlation matrix $\mathbf{\Xi}^{naive} = \mathbf{E}$, which is considered for comparative purposes. It will serve us to measure the improvement of the allocation strategies under the RMT estimators.

\section{Portfolio Theory}

The essential ingredients of portfolio theory are the expected return, risk, and asset allocation weights that optimize the risk/return trade-off.
Consider a universe of $p$ assets and denote by $s_{i,t}$ the asset price $i=1,\dots,p$ at time $t=1,\dots,n$. Thus, the logarithm return $r_{i,t}$ is defined as
 \begin{equation}
 r_{i,t} = \log\left(\frac{s_{i,t}}{s_{i,t-1}}\right)
 \end{equation}

Furthermore, the portfolio weight vector 
 \begin{equation}
  \mathbf{w} = \{w_1,\dots,w_p\},
 \end{equation}
represents the amount of money invested in the asset $i$ and can be positive or negative. In the former case, the investor is the owner of the stock and holds it expecting to sell it at a higher price. The strategy is known as having a \emph{long position}. A negative weight means the investor sells a borrowed stock expecting to buy it later at a lower price, known as \emph{short position}.

The expected return of the portfolio is defined in terms of the vector of expected profits$\mathbf{g}$ 
\begin{equation}
\mathcal{G} = \mathbb{E}(\mathbf{w'r}) = w'\mathbb{E}(r) = \mathbf{w'g},
\label{expected_profit}
\end{equation}
and the associated portfolio risk is defined as a function of covariance matrix $\mathbf{\Sigma}$ of the returns
\begin{equation}
    {R}^2 = \mathbf{w'\Sigma w}.
\end{equation}

\subsection{Markowitz}

The mean-variance allocation strategy of Markowitz~\cite{roncalli2013introduction} proposes to solve the following quadratic optimization problem to minimize the portfolio risk at a given level of expected return
\begin{equation} 
\underset{\mathbf{w}\in\mathbb{R}^p}{min} \frac{1}{2} \mathbf{w'\Sigma w} \quad - \gamma\mathbf{w}'\mathbf{g} \geq \mathcal{G},
\label{mean_variance}
\end{equation}
where $\gamma$ is interpreted as the risk-aversion parameter.
Hence, without considering any constraint on the weights and only assuming that the inverse of $\mathbf{\Sigma}$ exists, the optimal portfolio is found to be
 \begin{equation}
  \hat{\mathbf{w}} = \gamma \mathbf{\Sigma}^{-1} \mathbf{g}
  \label{optimal}
 \end{equation}

The minimum risk associated with this optimal weight solution is given by
\begin{equation}
   \mathcal{R}^2_{true} = \frac{\mathcal{G}^2}{\mathbf{g' \Sigma^{-1} g}}
   \label{7.5}
  \end{equation}
However,  the optimal solution is not achievable in any real situation, given that the covariance matrix $\mathbf{\Sigma}$ is unknown \emph{a priori}. 
Nevertheless,  it is possible to compute the in-sample risk $\mathcal{R}^2_{in}$ and the out-sample risk $\mathcal{R}^2_{out}$ using the expressions 
\begin{equation}
\begin{split}
& \mathcal{R}^2_{in} = \frac{\mathcal{G}^2}{\mathbf{g' E^{-1}_{in} g}},\\
& \mathcal{R}^2_{out} =   \frac{\mathcal{G}^2 \mathbf{g}'\mathbf{E^{-1}_{in}}\mathbf{E_{out}}\mathbf{E^{-1}_{in}}\mathbf{g}}{(\mathbf{g}'\mathbf{E^{-1}_{in}}\mathbf{g})^2},
\label{Risk_sample}
\end{split}
\end{equation}
where $\mathbf{E_{in}}$ and $\mathbf{E_{out}}$ are the historical and realized sample covariance matrices, respectively. 

An important result in the high-dimensional regime~($p,n \rightarrow \infty$, $q=p/n\rightarrow(0,1)$) derived from RMT suggest the following inequality~\cite{bun2017cleaning,potters2020first}
\begin{equation}
\frac{\mathcal{R}_{in}^{2}}{1-q} = \mathcal{R}_{true}^2 = (1-q)\mathcal{R}_{out}^2
\end{equation}
where it is assumed that the direction of $\mathbf{g}$ is independent of $\mathbf{\Sigma}$ or $\mathbf{E}$, such that $\mathbf{g}$ is normalized as $\mathbf{g'g} = p$. 
This inequality makes evident the fact that there is a bias in the estimate of the true optimal frontier $\mathcal{R}^2_{out}$ as the parameter $q$ approaches 1, that is, as the number of assets is of the order of the number of transaction times. Therefore, special attention should be paid to this regime and consider the covariance estimators presented in the previous section.

\subsection{NCO}

\begin{figure}[!ht]
\centering
\begin{tikzpicture}[node distance=1.25cm]

\node (start) [startstop] {Input:  $\mathbf{\Xi}$};
\node (in1) [io, below of=start] {(1) $\mathbf{D}$};
\node (in2) [io, below of=in1] {(2) MST($\mathbf{D}$)};
\node (in3) [io, below of=in2] {(3) $\mathbf{Lv} = \lambda \mathbf{v}$};
\node (in4) [io, below of=in3] {(4) $\max\limits_{k}|\lambda_{k} - \lambda_{k+1}|$};
\node (in5) [io, below of=in4] {(5) SC(k)};
\node (in6) [io, below of=in5] {(6) $\mathbf{\Xi}_{sc}$};
\node (in7) [io, below of=in6] {(7) $\mathbf{w}_{intra}$};
\node (in8) [io, below of=in7] {(8) $\mathbf{w}_{inter}$};
\node (stop) [startstop, below of=in8] {Output: $\mathbf{w} = \mathbf{w}_{intra}\mathbf{w}_{inter}$};

\draw [arrow] (start) -- (in1);
\draw [arrow] (in1) -- (in2);
\draw [arrow] (in2) -- (in3);
\draw [arrow] (in3) -- (in4);
\draw [arrow] (in4) -- (in5);
\draw [arrow] (in5) -- (in6);
\draw [arrow] (in6) -- (in7);
\draw [arrow] (in7) -- (in8);
\draw [arrow] (in8) -- (stop);

\end{tikzpicture}
\caption{Flowchart representing the steps of the NCO allocation strategy. The input is the covariance matrix $\mathbf{\Xi}$, and the output the vector of optimal weights $\mathbf{w}$.}
\label{flowchart}
\end{figure}
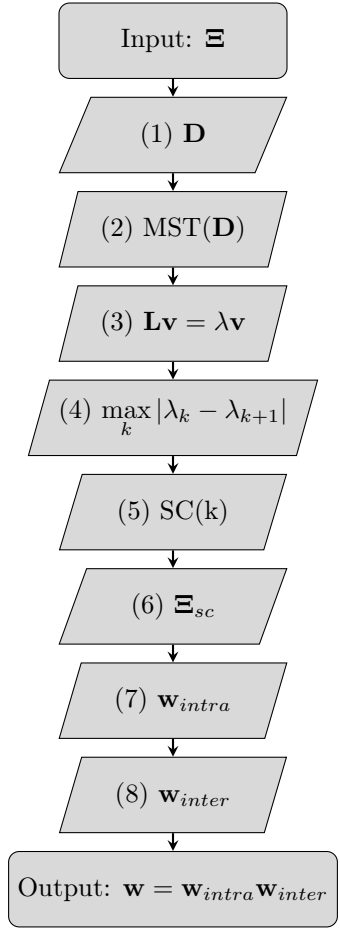

This asset allocation proposal consists of a series of steps represented as a flowchart in Fig.~\ref{flowchart}.
The input is the correlation matrix $\mathbf{\Xi}\in\{\mathbf{\Xi}^{naive},\mathbf{\Xi}^{TW},\mathbf{\Xi}^{linear}\}$. In step 1 it is computed the dissimilarity matrix $\mathbf{D}$ of $\mathbf{\Xi}$ by
\begin{equation}
    D_{ij} = \sqrt{\frac{1}{2}(1-\Xi_{ij})} 
\end{equation}
where $i,j=1,\dots,p$ are the indices of the matrix elements of $\mathbf{D}$ and $\mathbf{\Xi}$. Next, it is applied the MST algorithm over $\mathbf{D}$ to obtain a weighted undirected graph without cycles $G_{MST}$ that minimizes the total edge weights~(step 2). 
In step 3, the weights of the graph $G_{MST}$ are inverted to $1/w$ to interpret high values as strongly connected and low values as weakly connected.
Then, the normalized Laplacian of $G_{MST}$ is calculated by
\begin{equation}
    \mathbf{L} = \mathbf{B}^{-1/2}(\mathbf{B-W})\mathbf{B}^{-1/2},
\end{equation}
where $\mathbf{B}$ is a diagonal matrix with degrees $b_1,\dots, b_p$ on the diagonal, and $\mathbf{W}$ is the weighted matrix of $G_{MST}$. 
Then, the pairs ($\lambda, \mathbf{v}$) of eigenvalues and eigenvectors are obtained through the eigendecomposition of $\mathbf{L}$.
Next, the optimal number of clusters is estimated using the eigengap or \emph{spectral gap} procedure~\cite{von2007tutorial}~(step 4). It keeps the index $k$ that maximizes the absolute difference of eigenvalues $|\lambda_{k}-\lambda_{k+1}|$. The spectral clustering is now applied to the matrix formed by the $k$ eigenvectors employing the standard k-means algorithm~(step 5). 
Thus, the covariance matrix elements are rearranged following the block structure of the $k$ formed clusters to get the sorted matrix $\mathbf{\Xi}_{sc}$~(step 6). Hence, the intracluster weights $\mathbf{w}_{intra}$ are computed on each of the $k$ correlation blocks through the mean-variance expression of Eq.~\ref{mean_variance}~(step 7), and the intercluster weights $\mathbf{w}_{inter}$  are computed analogously between the $k$ clusters~(step 8). Finally, the output is the allocation weights $\mathbf{w}$ given by the product of $\mathbf{w}_{intra}$ and $\mathbf{w}_{inter}$. 
In this way, the Markowitz curse is transformed into a well-behaved problem because the correlation matrix is more similar to a diagonal matrix~\cite{lopez2016robust}.

\section{Data}

We consider the close prices of companies listed on the Mexican Stock Exchange from 2017-12-29 to 2022-07-01 at a weekly frequency.
The data were collected from \url{https://finance.yahoo.com/}, and the returns are computed through the logarithm differences of the closing prices.
Only markets with less than 10\% missing days were included.
Thus, we study a set of $p=28$ return times series of length $T=232$ transaction weeks imputed by linear interpolation.

\begin{table}[!ht]
    \centering
    \small
\begin{tabular}{|l|l|l|}
\hline
Ticker & Name \\ \hline
AC.MX & Arca Continental, S.A.B. de C.V.  \\
AEROMEX.MX & Grupo Aeroméxico, S.A.B. de C.V.  \\
AGUA.MX & Grupo Rotoplas S.A.B. de C.V. \\
ALFA.MX & Alfa S.A.B. de C.V. \\
ALSEA.MX & Alsea, S.A.B. de C.V.  \\
ARA.MX & Consorcio ARA, S. A. B. de C. V. \\
BBVA.MX & Banco Bilbao Vizcaya Argentaria, S.A. \\
CETETRC.MX & iShares S\&P/VALMER Mexico CETETRAC \\
CORPTRC.MX & iShares Mexico Corporate Bond TRAC \\
CUERVO.MX & Becle, S.A.B. de C.V.  \\
ELEKTRA.MX & Grupo Elektra, S.A.B. de C.V.  \\
GCC.MX & GCC, S.A.B. de C.V. \\
GENTERA.MX & Gentera, S.A.B. de C.V. \\
GMXT.MX & GMéxico Transportes, S.A.B. de C.V. \\
HCITY.MX & Hoteles City Express, S.A.B. de C.V.  \\
HERDEZ.MX & Grupo Herdez, S.A.B. de C.V.  \\
HOMEX.MX & Desarrolladora Homex, S.A.B. de C.V.  \\
HOTEL.MX & Grupo Hotelero Santa Fe, S.A.B. de C.V. \\
IVVPESO.MX & iShares S\&P 500 Peso Hedged TRAC \\
M10TRAC.MX & iShares S\&P/VALMER Mexico M10TRAC \\
NAFTRAC.MX & iShares NAFTRAC  \\
ORBIA.MX & Orbia Advance Corporation, S.A.B. de C.V. \\
PE\&OLES.MX & Industrias Peñoles, S.A.B. de C.V. \\
PINFRA.MX & Promotora y Operadora de Infraestructura, S. A. B. de C. V.  \\
Q.MX & Quálitas Controladora, S.A.B. de C.V.  \\
UDITRAC.MX & iShares S\&P/VALMER Mexico UDITRAC \\
VESTA.MX & Corporación Inmobiliaria Vesta, S.A.B. de C.V. \\
WALMEX.MX & Wal-Mart de México, S.A.B. de C.V.  \\
\hline
\end{tabular}
    \caption{Ticker and name of the considered $p=28$ financial instruments  trading on the BMV.}
    \label{Table1}
\end{table}

\begin{figure}[!ht]
 \centering
 \includegraphics[width=0.65\linewidth]{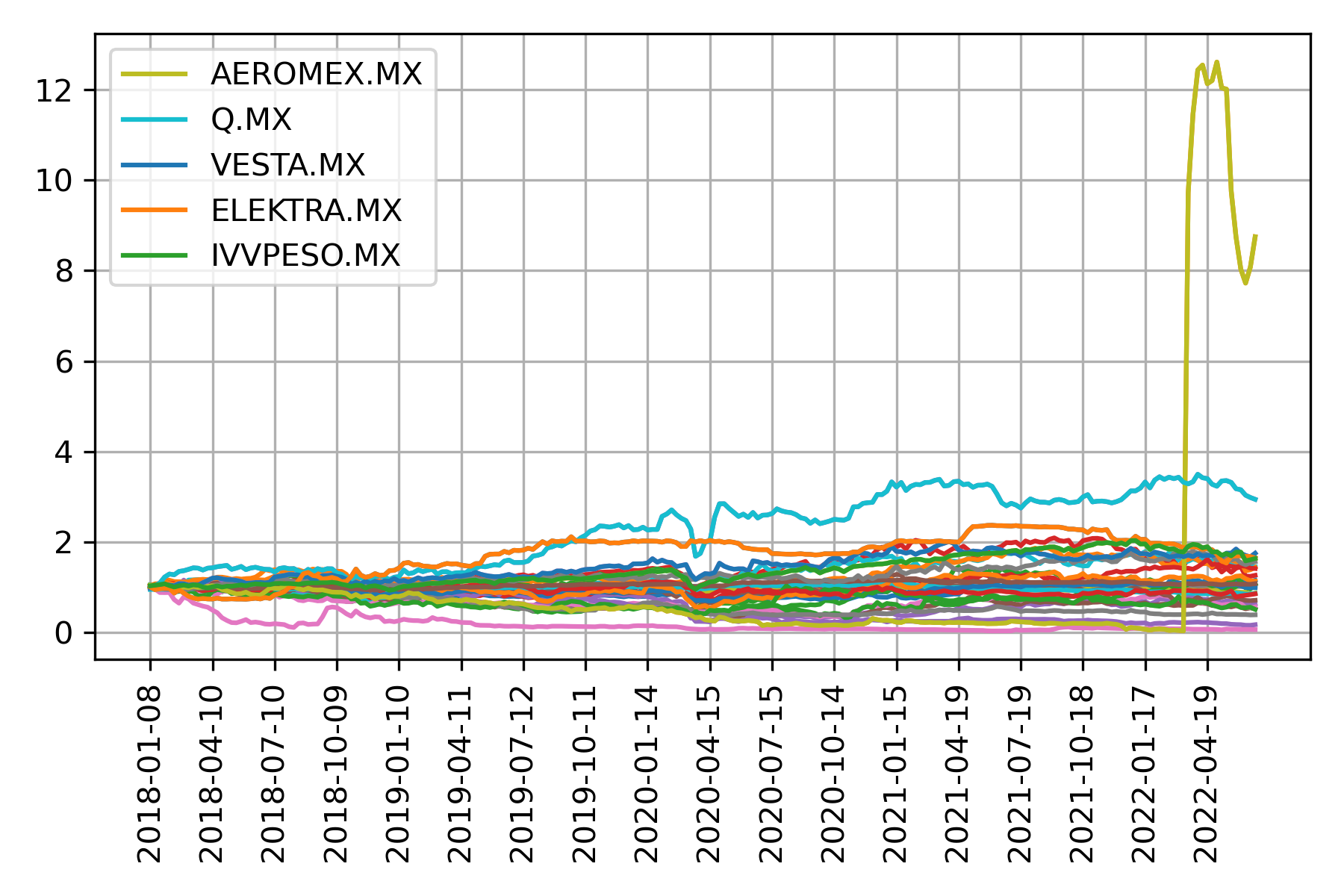}
 \caption{Cumulative returns. The top five companies with the best performances stand out on the legend. The starting values are normalized to one unit.}
 \label{Fig1}
\end{figure}

Fig.~\ref{Fig1} shows the cumulative returns of the companies under study.
It can be noticed that AEROMEX apparently presents the best performance, but the increment is a consequence of the \emph{inverse split} of the asset during March 2022, where the conversion factor was one new share for every 5,000,000 (five million) shares then existing~\footnote{\url{https://www.bmv.com.mx/docs-pub/eventemi/eventemi_1170108_1.pdf}}

Next, subsamples of $n=2p$ observations shifted by $\Delta t=1$ week were created to obtain a total of $m=121$ windows of weekly logarithmic returns.
We concentrate on correlation instead of volatilities. Then, the returns of each data window are standardized by three steps. First, remove the sample mean $\mu_i$ of each asset. Second,  normalize the returns by the cross-sectional weekly volatility $\hat{\sigma}_{i,t} = \sqrt{\sum_j r_{j,t}}$, to obtain an estimation $\hat{r}_{i,t} = r_{it}/\hat{\sigma}_{i,t}$. Third, normalize the returns by the sample volatility $\sigma_i$:  $\tilde{r}_{i,t} = \hat{r}_{i,t}/\sigma_i$. Particularly, the second step removes a substantial amount of non-stationarity in the volatilities. Hence, the returns are now stationary at first order, and we have a well-behaved dataset to apply our proposed methodology.

\section{Results}

We compute the  $\mathbf{E}_t$, $\mathbf{\Xi}^{linear}_t$ and $\mathbf{\Xi}^{TW}_t$ estimators to each data window $t=1,\dots, m$, where the estimated covariance matrices were transformed into correlation matrices for comparative purposes. Thus, from now on we will refer to correlation matrices instead of covariance matrices. Nevertheless, the portfolio theory can be applied indistinctly. 

\begin{figure}[!ht]
 \centering
 \includegraphics[width=0.45\linewidth]{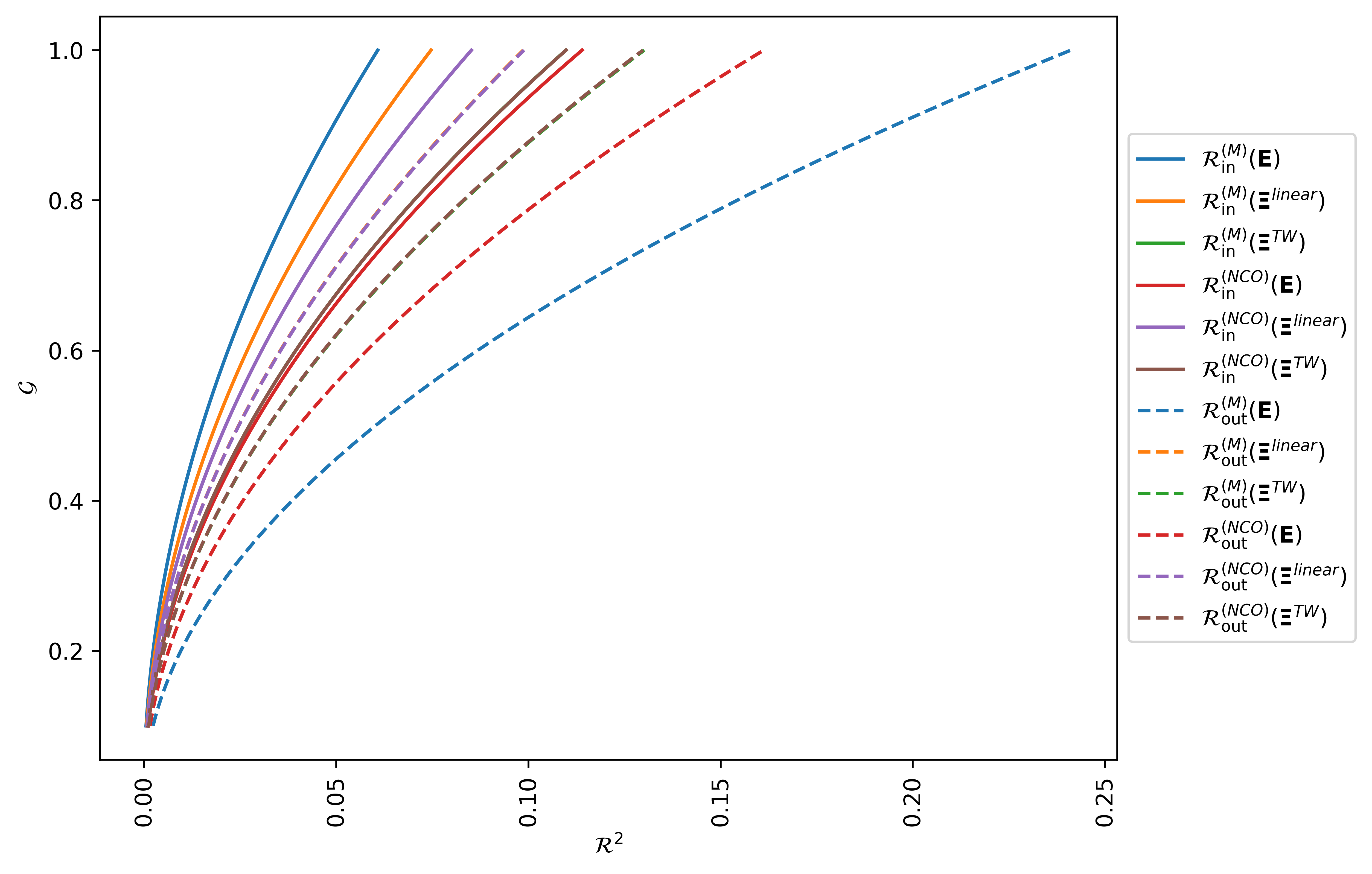}
 \includegraphics[width=0.45\linewidth]{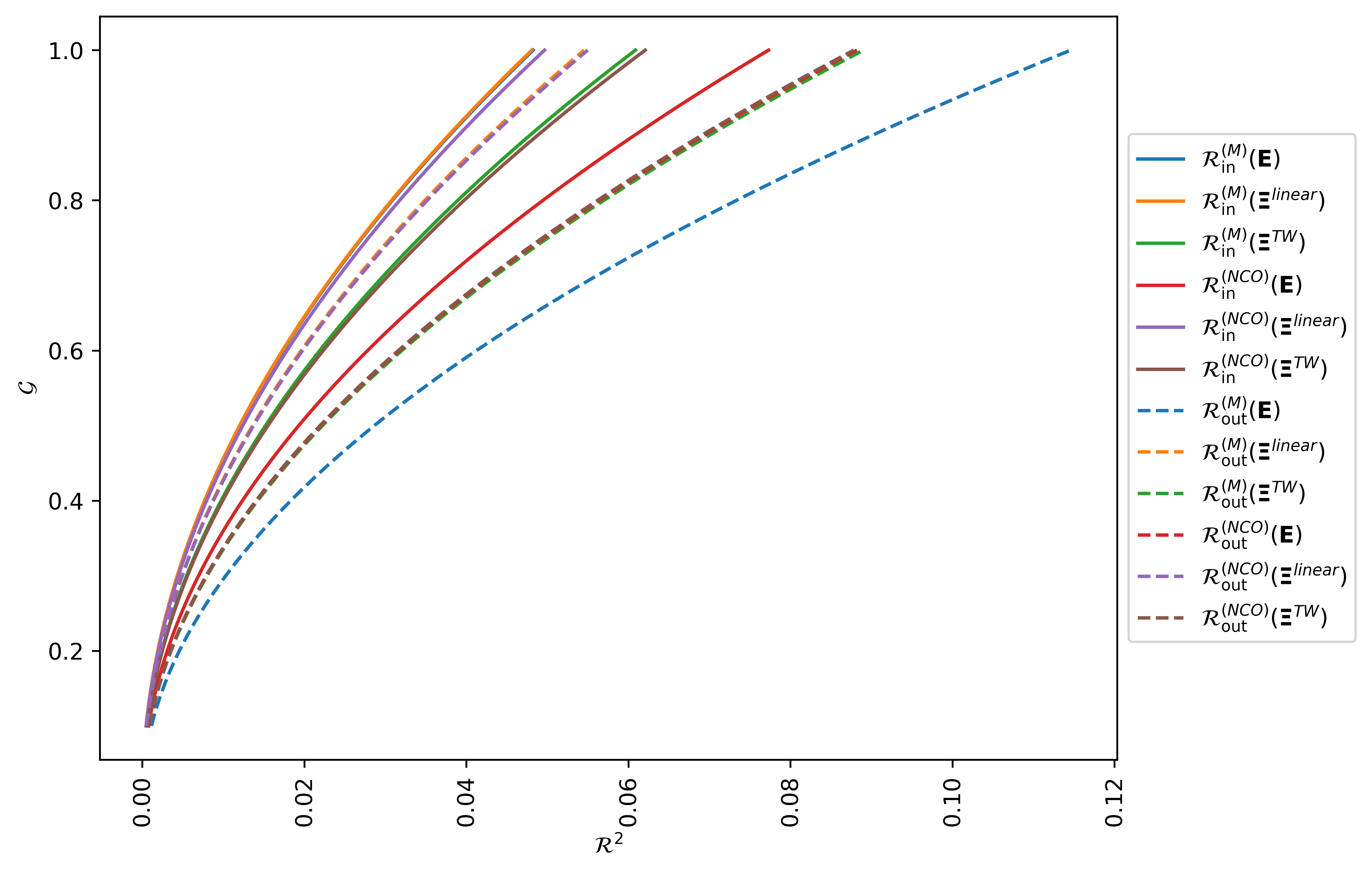}
 \caption{Efficient frontier of the Markowitz and NCO allocation methodologies under the different estimation strategies of the correlation matrices.  (a) $t=1$. (b) $t=m$. In all cases is considered a minimum variance portfolio scenario~($\mathbf{g}=1$) and the dimensional factor  $q=1/2$. The superscript refers to the associated allocation methodology: M (Markowitz) or NCO.}
 \label{Fig2}
\end{figure}

Fig.~\ref{Fig2} shows  the in-sample and out-sample efficient frontier of the Markowitz and NCO allocation methodologies for the first and last sample periods under the different estimation strategies of the correlation matrices. 
Here, the in-sample and out-sample frontiers are built by the corresponding correlation matrix estimations at $t$ (in-sample) and $t+1$(out-sample), and assuming a minimum variance portfolio scenario~($\mathbf{g}=1$). 
In general, it can be observed that the estimators $\mathbf{E}_t$ and $\mathbf{\Xi}^{linear}_t$ reduce the gap between the out-sample and in-sample efficient frontier for both asset allocation methodologies.

\begin{figure}[!ht]
 \centering
 \includegraphics[width=0.65\linewidth]{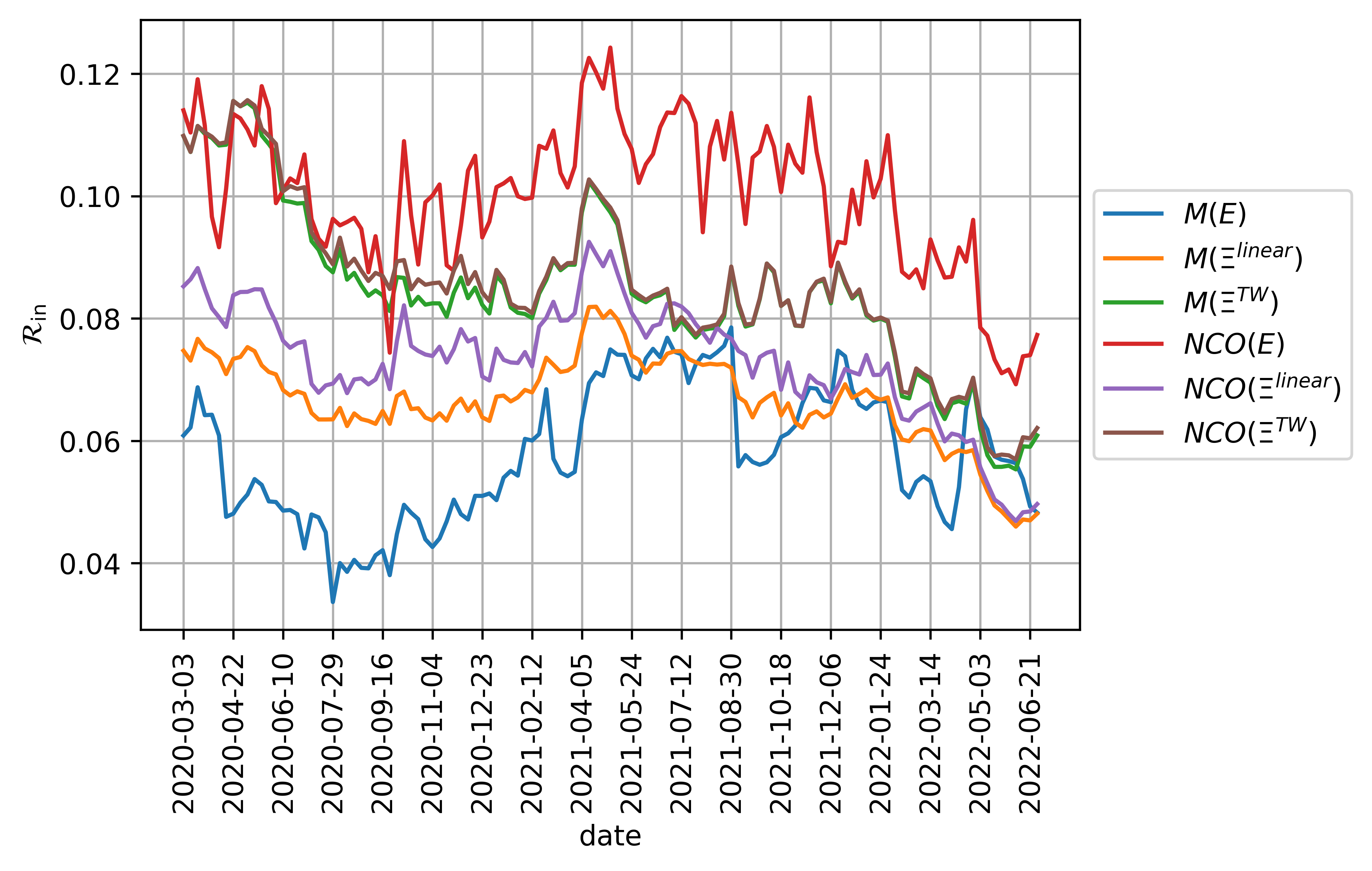}
 \includegraphics[width=0.65\linewidth]{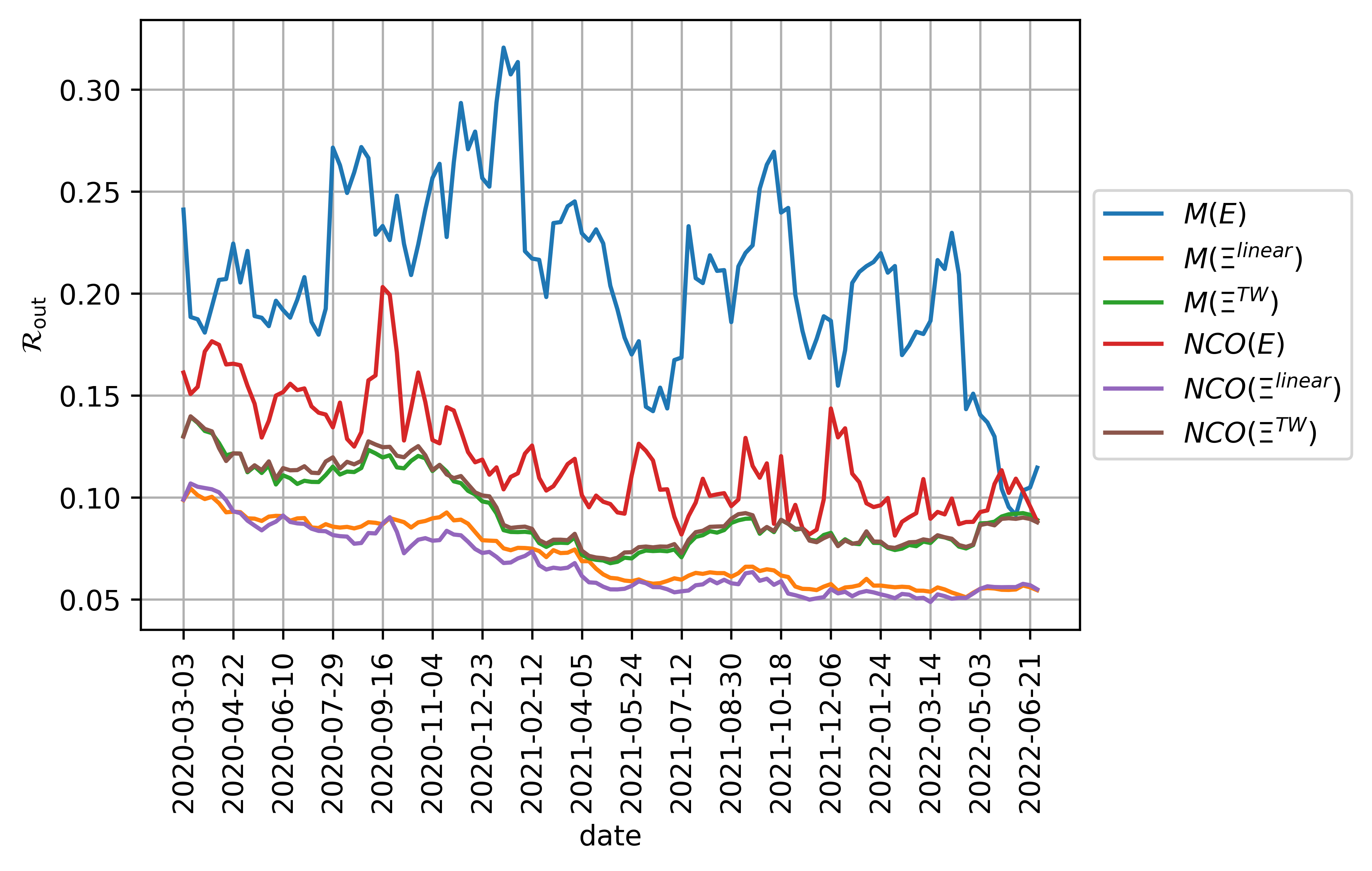}
 \includegraphics[width=0.65\linewidth]{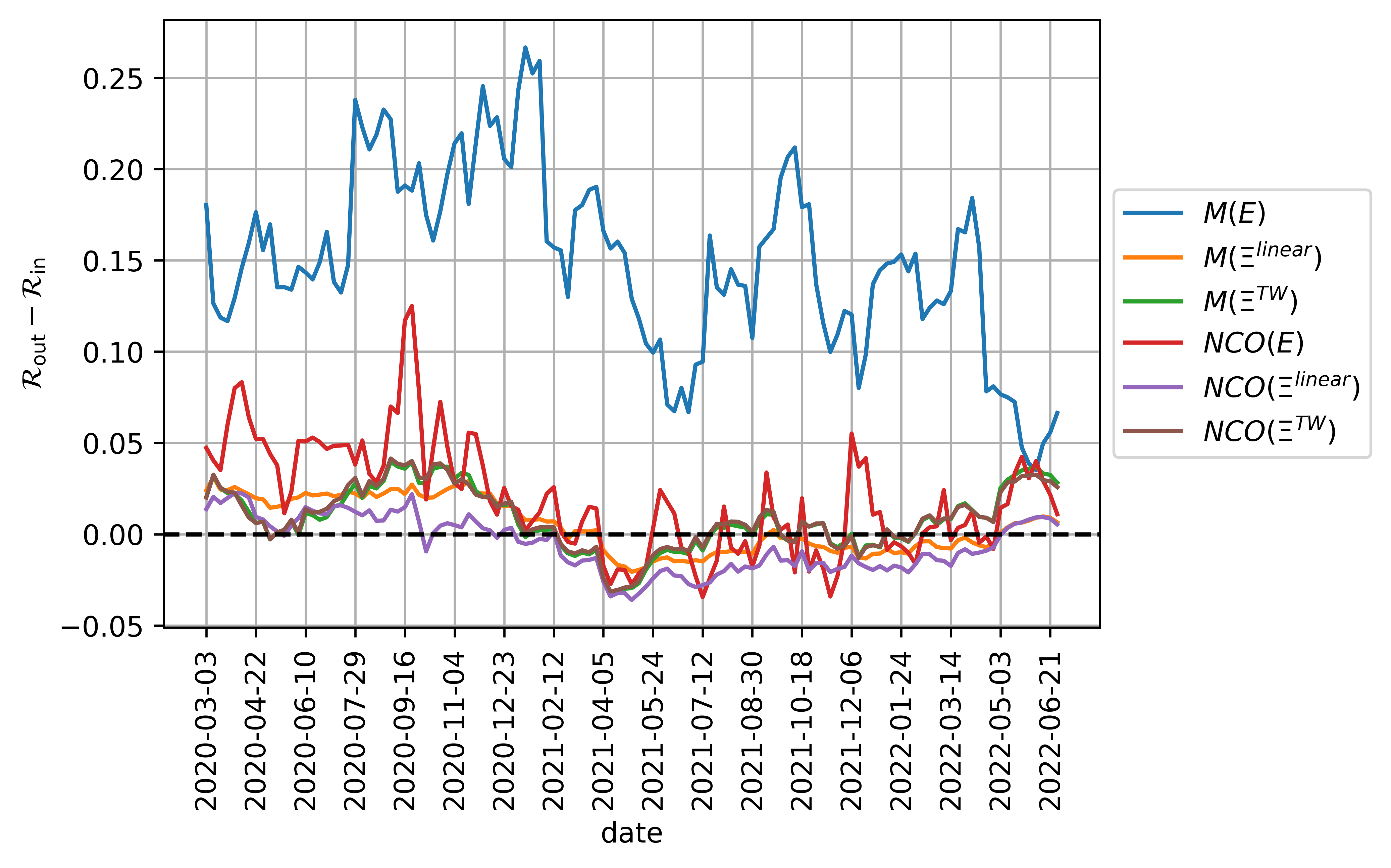}
    \caption{Risk dynamics for $t=1, \dots, m$. Top:  in-sample risk. Middle: out-sample risk. Bottom: the difference between the out-sample and in-sample risk. In all cases is considered a minimum variance portfolio scenario~($\mathbf{g}=1$) at a fixed level $\mathcal{G}=1$ and dimensional factor $q=1/2$. The label refers to the associated allocation methodology and estimator.}
 \label{Fig3}
\end{figure}

Next, the in-sample and out-sample risk of a minimum variance portfolio~(see Eq.~\ref{Risk_sample}) was calculated for $\mathbf{E}_t$ in each frame $t=1,\dots, m$; setting $\mathcal{G}=1$. 
Fig.~\ref{Fig3} shows the risk dynamics of the minimum variance portfolio through time under each allocation methodology and estimator of the correlation matrix.
When applying the estimators of $\mathbf{E}$, the in-sample risk increases under the Markowitz methodology, while it decreases with NCO (see Fig.~\ref{Fig3}a). Then, the NCO methodology on the raw empirical matrix is considered the best strategy to avoid underestimating the in-sample risk.
The results show that Markowitz and NCO methodologies reduce the overestimation of the out-sample risk under the linear and TW estimators (see Fig.~\ref{Fig3}b). Nevertheless, the combination of NCO with the linear estimator is the best strategy to deal with the out-sample risk overestimation.
The most exciting quantity to minimize is the difference or gap between the out-of-sample and in-sample risks. It can be seen that this gap is significantly minimized when we use the linear estimator, where in some periods, the optimal is obtained by combining it with the Markowitz methodology and in others with the NCO~(see Fig.~\ref{Fig3}a).

\begin{figure}[!ht]
 \centering
 \includegraphics[width=0.3\linewidth]{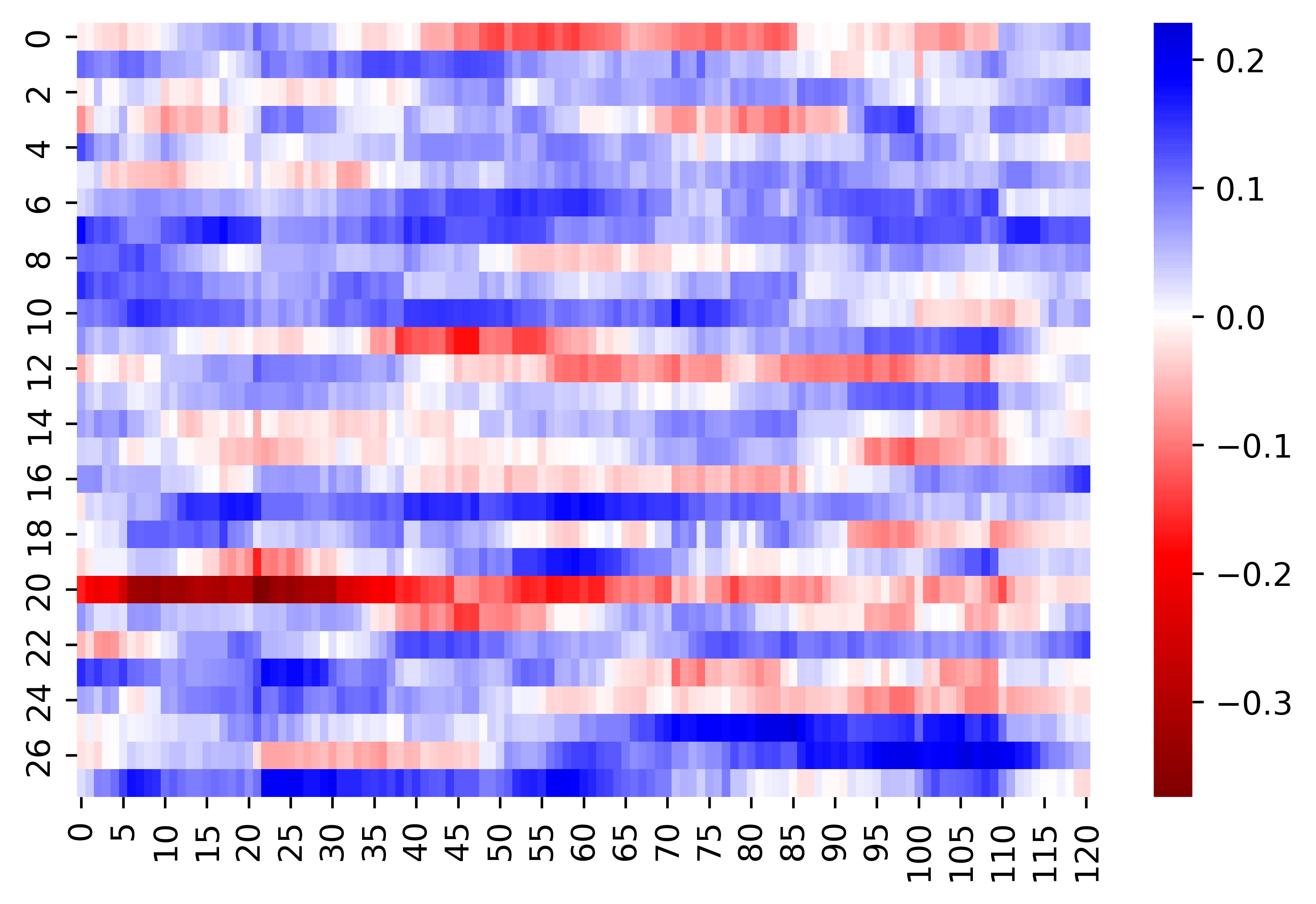}
 \includegraphics[width=0.3\linewidth]{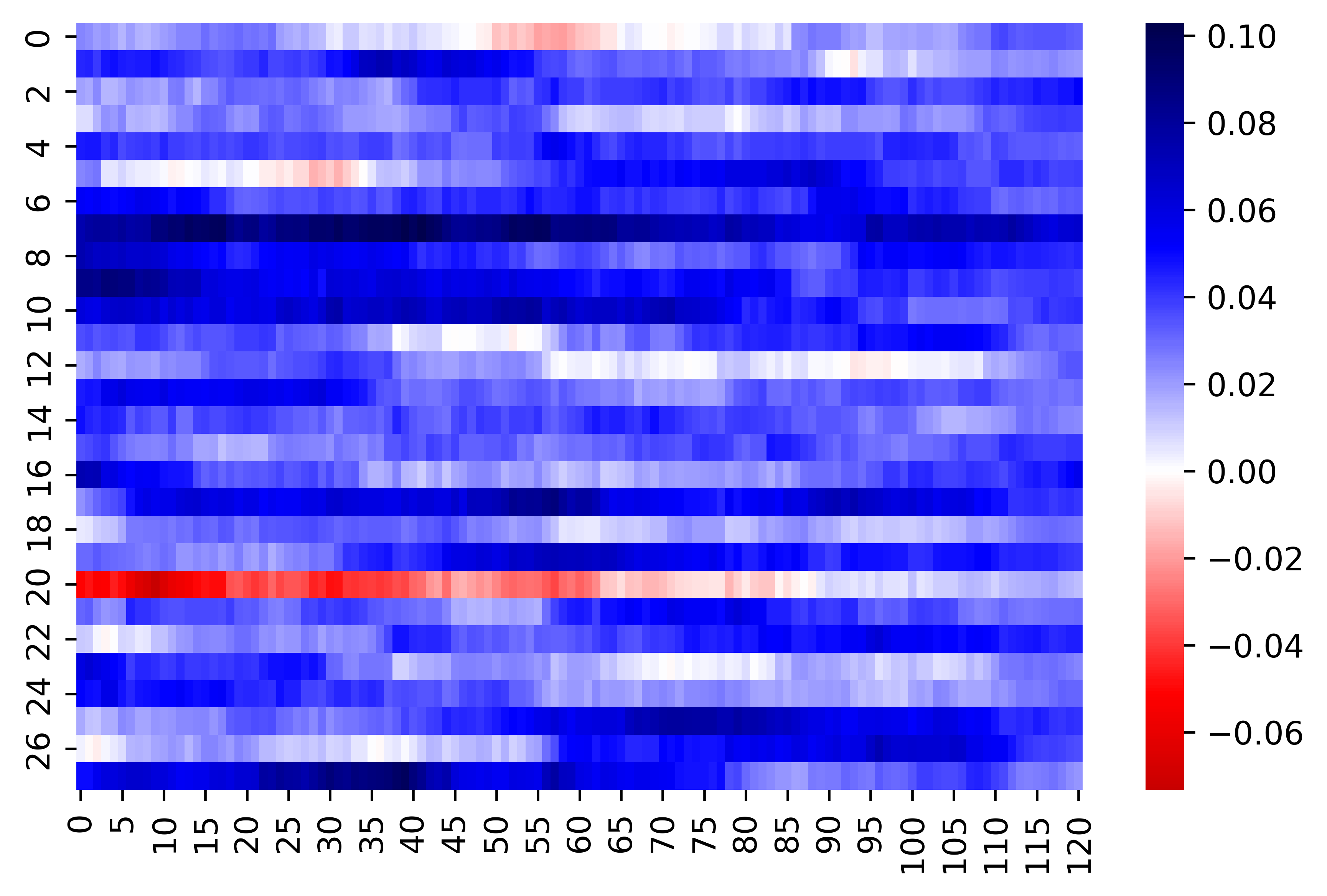}
 \includegraphics[width=0.3\linewidth]{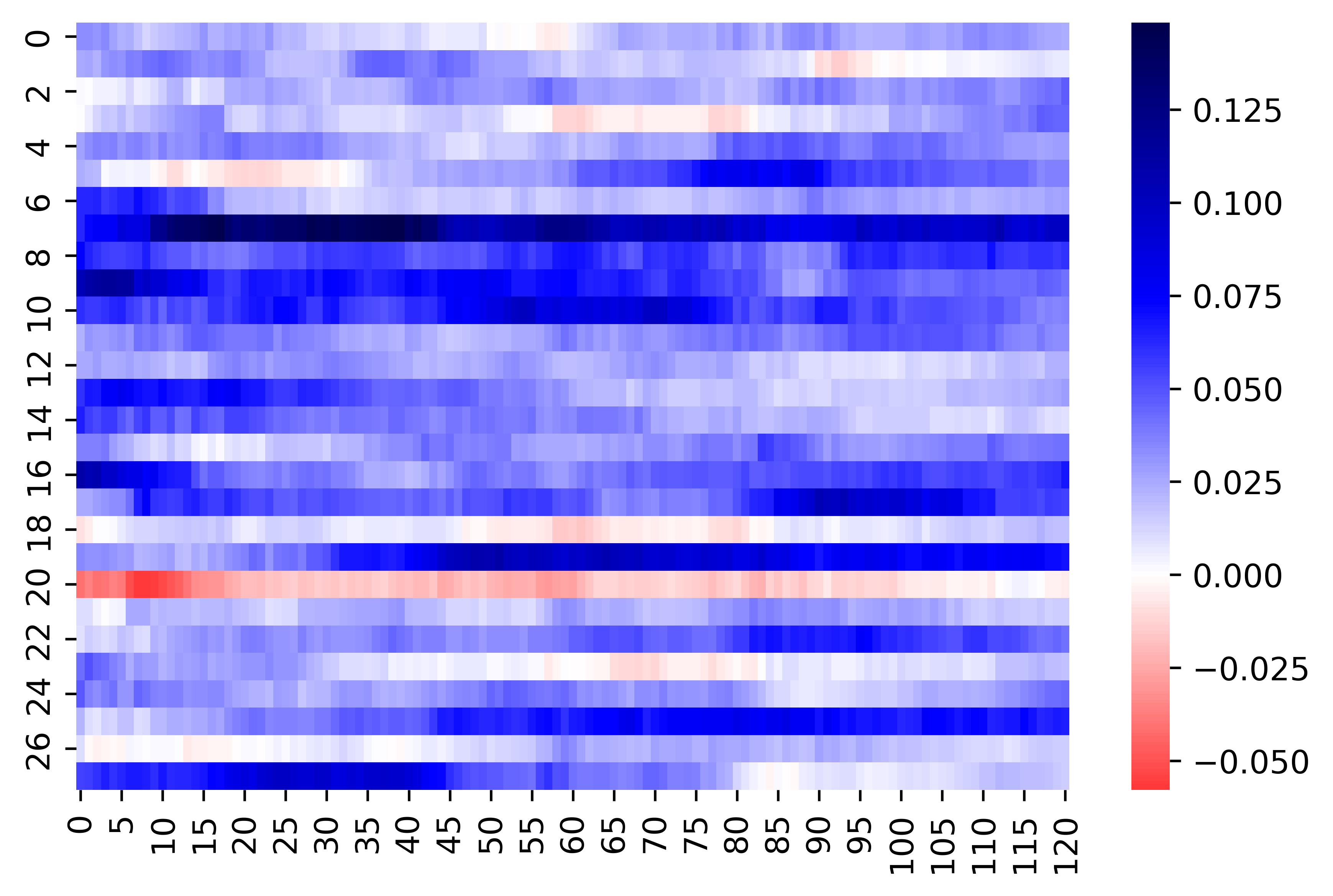}\\
 \includegraphics[width=0.3\linewidth]{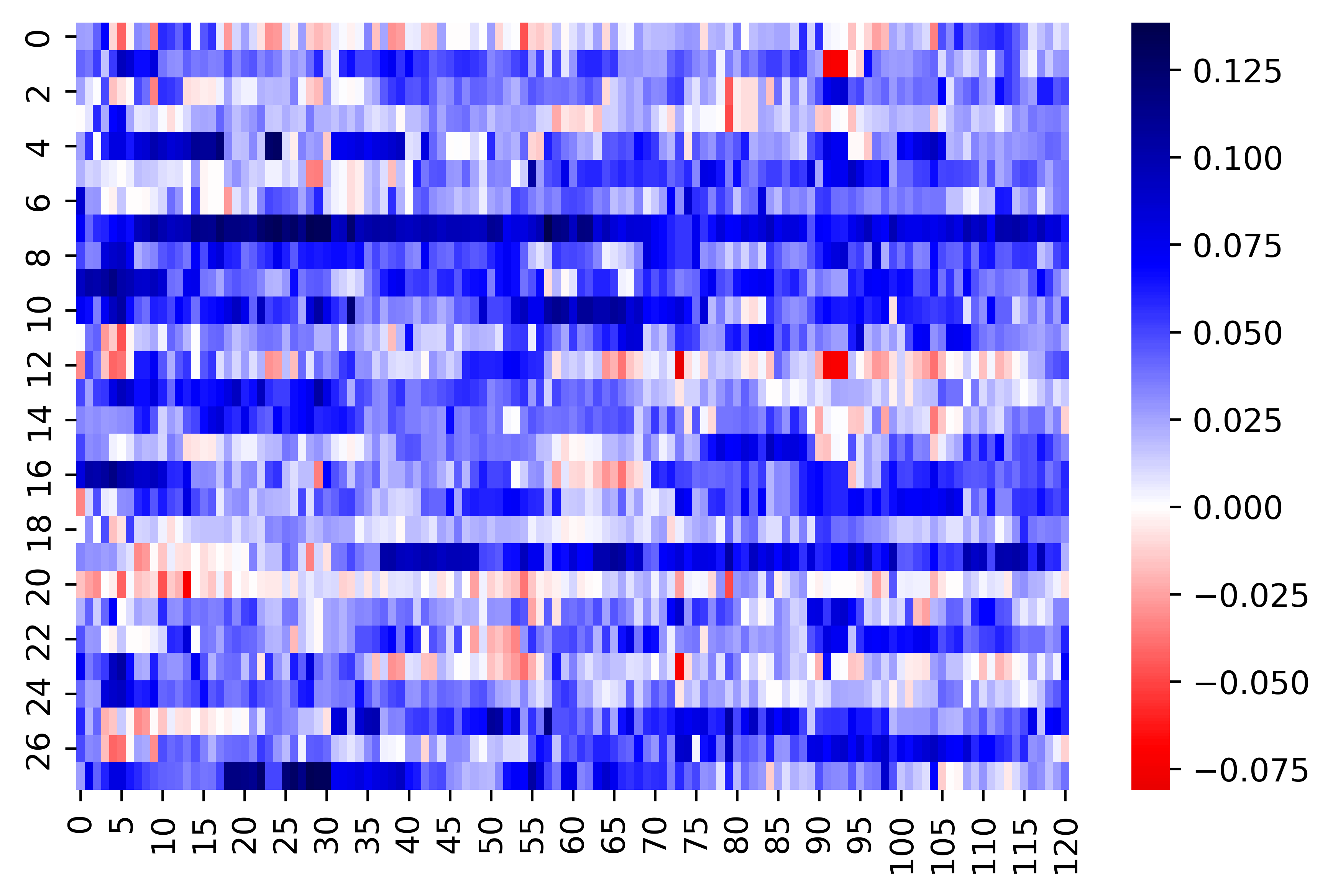}
 \includegraphics[width=0.3\linewidth]{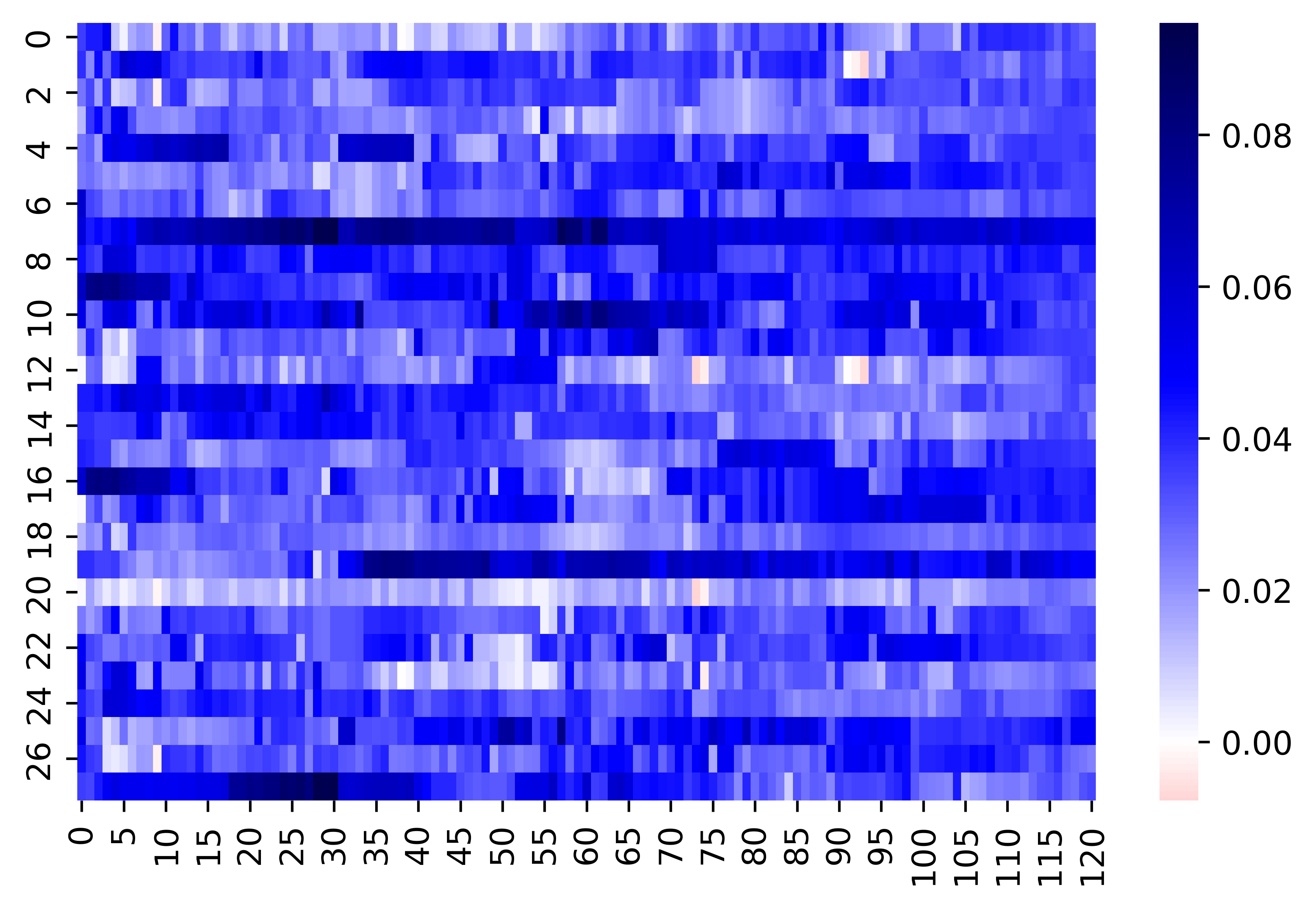}
 \includegraphics[width=0.3\linewidth]{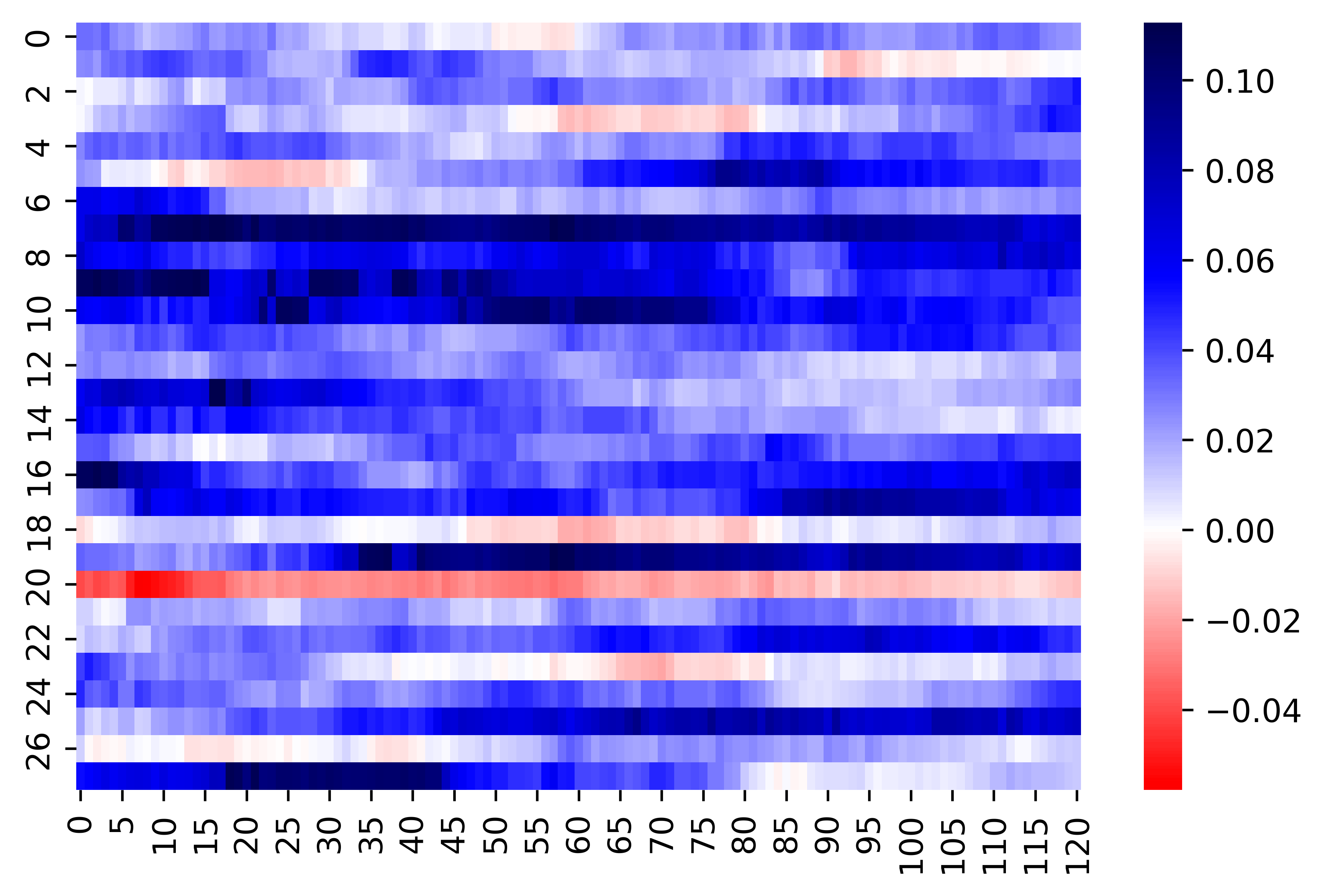}
    \caption{Allocation c each allocation methodology and estimator of the correlation matrix. 
    Top left: Markowitz with $\mathbf{E}$.
    Top middle: Markowitz with $\mathbf{\Xi}^{linear}$.
    Top right: Markowitz with $\mathbf{\Xi}^{TW}$.
    Bottom left: NCO with $\mathbf{E}$.
    Bottom middle: NCO with $\mathbf{\Xi}^{linear}$.
    Bottom right: NCO with $\mathbf{\Xi}^{TW}$.
    The x-axis refers to the period and the y-axis to the market~(listed in the same order as Table~\ref{Table1}). In all cases is considered a minimum variance portfolio scenario~($\mathbf{g}=1$) at a fixed level $\mathcal{G}=1$ and dimensional factor $q=1/2$.}
 \label{Fig4}
\end{figure}

Figure \ref{Fig4} shows the portfolio composition through time ($t=1,\dots, m$) for each allocation methodology and estimator of the correlation matrix under the setting $\mathbf{g}=\mathbf{1}$ and $\mathcal{G}=1$.
It is observed that there are optimal strategies where the weights are outside the domain [0,1], which implies that a short sale strategy must be applied, that is, borrowing shares to return them at the market price later. These strategies are extremely risky and only practiced by experienced investors.
Interestingly, for the same level of expected gain, the combination of the NCO methodology with the linear filter is the only one that gives us non-negative weights for all periods. While if we do not apply the estimators, the number of negative weights increases considerably in the NCO and Markowitz methodologies. The latter is the one that shows greater magnitudes, both positive and negative.

\begin{figure}[!ht]
 \centering
 \includegraphics[width=0.65\linewidth]{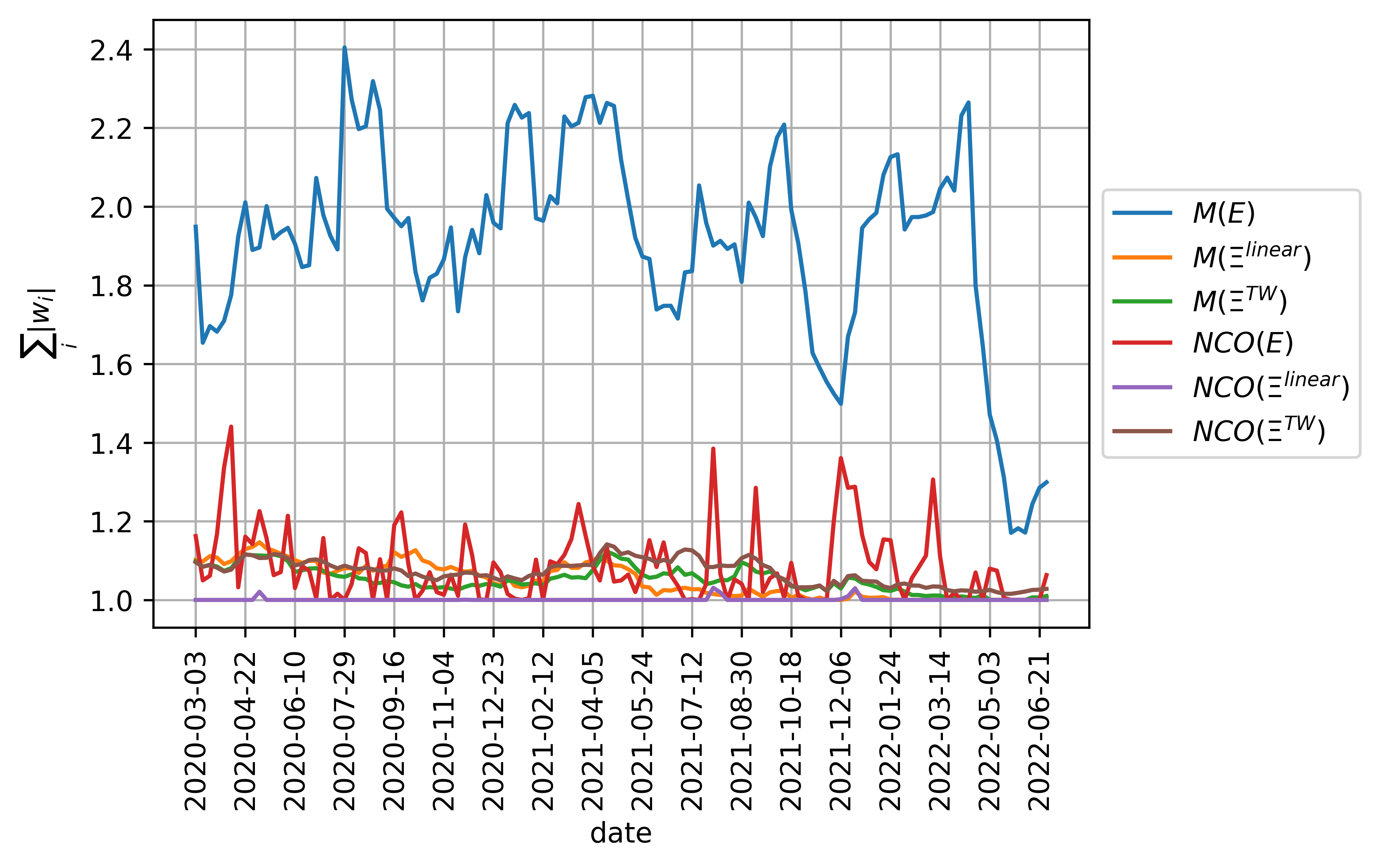}
 \caption{Absolute sum of weights for the different allocation strategies. The label refers to the associated allocation methodology and correlation matrix estimator. In all cases is considered a minimum variance portfolio scenario~($\mathbf{g}=1$) at a fixed level $\mathcal{G}=1$ and dimensional factor $q=1/2$.}
 \label{Fig5}
\end{figure}

Fig.~\ref{Fig5} shows the absolute value sum of the weights associated with the minimum variance portfolios built under each allocation and estimator strategy considering the same settings as Fig~\ref{Fig4}.
The linear and TW estimators clearly reduce the absolute sum of the weights. Notably, the NCO strategy always keeps the absolute sum of the weights very close to one. This behavior translates into greater diversification without exposure to short-selling strategies.
It is interesting to note that the instrument CETETRC.MX (row 7) is the one with the highest proportion of capital assigned in most periods and scenarios. Therefore, a preference for Mexican government bonds as an investment strategy is reflected in the solutions.

Finally, table~\ref{Table2} shows the performance metrics averaged over all periods. The Mean Square Error (MSE) and Mean Absolute Error (MAE) were applied to the difference between the in-sample and out-sample risks. It can be seen that the best results are obtained for the Markowitz methodology under the linear estimator. Also, the NCO methodology yields optimal values with the same linear estimator. In the case of asset allocation, the amount called the Mean Sum of Absolute Weights (MSAW) was calculated, which gives us an idea of how the weights were distributed over time. It can be verified what was seen in figure~\ref{Fig5}, that is, the absolute sum is minimized with NCO under the linear estimator during the study period, obtaining an average value practically equal to one.
\begin{table}[!ht]
 \centering
 \caption{Performance Metrics. The first column denotes the allocation strategy. The second and third columns show the Mean Square Error~(MSE) and Mean Absolute Error~(MAE) of the difference between the in-sample and out-sample risk, respectively. The last column shows the Mean Sum of Absolute Weights~(MSAW). The average is computed over the m=121 data window setting on a minimum variance portfolio scenario~($\mathbf{g}=1$) at a fixed level $\mathcal{G}=1$ and dimensional factor $q=1/2$.}
    \begin{tabular}{|l|l|l|l|}
    \hline
    Case & MSE & MAE & MSAW \\ \hline
    $M(E)$ & 0.025005 & 0.149937 & 1.909886 \\ \hline
    $M(\Xi^{linear})$ & {\bf 0.000238} & {\bf 0.013284} & 1.049836 \\ \hline
    $M(\Xi^{TW})$ & 0.000360 & 0.014970 & 1.050949 \\ \hline
    $NCO(E)$ & 0.001302 & 0.027702 & 1.083757 \\ \hline
    $NCO(\Xi^{linear})$ & 0.000255 & 0.013803 & {\bf 1.000924} \\ \hline
    $NCO(\Xi^{TW})$ & 0.000353 & 0.014983 & 1.070408 \\
    \hline
    \end{tabular}
    \label{Table2}
\end{table}

\section{Conclusion}

The asset allocation problem suffers from significant instability from the classical Markowitz perspective due to highly correlated assets increase the condition number. A highly correlated portfolio is common during systemic contagion events since they lead to co-movements within the financial markets.
At the same time, we are subject to periods of structural changes in the financial markets derived from financial turmoil.
Then, it is preferable to consider a relatively small number of records to avoid the non-stationarity phenomena, but a high number of assets in order to diversify as much as possible the investment portfolio. 
As such,  we face situations where the number of assets is the order of the number of observations. Afterward, the instability derived from Markowitz's curse is exacerbated in these high-dimensional scenarios.
Fortunately, the estimators of the covariance matrix from RMT allow us to deal with the bias or noise coming from high dimensionality. Furthermore, the NCO methodology helps us avoid Markowitz's curse. Therefore, combining both strategies is an effective solution to deal with several financial problems in the investment management.

It is crucial to control portfolio allocation weights for two fundamental reasons.
On the one hand, solutions with low values of the absolute sum of the weights are associated with diversified portfolios that help minimize the investment risk. 
On the other hand, positive weights mean that it is only required to allocate capital to assets owned by the investor.
A remarkable discovery is that when NCO is applied together with the linear estimator, diversified solutions with non-negative weights are obtained whose absolute values give us a sum close to one.
Further, control over capital allocations is carried out for the same level of expected profit. Therefore, it has been possible to maximize profit without applying extremely risky investment strategies. Even so, the Markowitz mean-variance model, in conjunction with the linear estimator, is the one that best controls the in-sample and out-sample risks in absolute numbers.

Surprisingly, both portfolio optimization methodologies have not been subject to the restriction that the weights need to be positive or sum to one.
Notwithstanding, we have obtained restricted solutions without the need to model them as a quadratic programming problem with linear constraints, which would increase the computational complexity and possibly the stability of the solutions.
Consequently, as future work, it is necessary to delve into the particular mechanism of the NCO algorithm and the RMT estimators that allow achieving this control over capital allocations.
It is intriguing to know in what sense spectral clustering and MST have been accomplices to stabilize the portfolio weights of the BVM.

\section*{Acknowledgements}
This research was funded by Consejo Nacional de Humanidades, Ciencias y Tecnologías~(CONAHCYT, Mexico) through fund FOSEC SEP-INVESTIGACION BASICA [A1-S-43514].

\printbibliography
% %\bibliographystyle{rmf-style}
% \bibliography{ref}

\end{document}